\documentclass[11pt]{article}
\usepackage{moriond,epsfig}

\def\Journal#1#2#3#4{{#1} {\bf #2}, #3 (#4)}

\def\PLB{{\em Phys. Lett.}  B}
\def\PRL{\em Phys. Rev. Lett.}
\def\PRD{{\em Phys. Rev.} D}
\def\ZPC{{\em Z. Phys.} C}

\def\be{\begin{equation}}
\def\ee{\end{equation}}
\def\bea{\begin{eqnarray}}
\def\eea{\end{eqnarray}}

\def\etal{{\it et al.}}
\def\kz{\mbox{$K^0$}}
\def\kzb{\mbox{$\overline{K^0}$}}
\def\pL{\mbox{${p\bar\Lambda}$}}
\def\LL{\mbox{${\Lambda\bar\Lambda}$}}
\def\ppbar{\mbox{${p\bar p}$}}

\def\ks{\mbox{$K^0_S$}}

\begin{document}
\vspace*{4cm}
\title{
Hadronic B Decays at CLEO 
}

\author{ J.E. Duboscq }

\address{
Wilson Laboratory, Cornell University, Ithaca NY 14853, USA
\\
XXXVIII Rencontres de Moriond on 
QCD and Hadronic Interactions, March 2003
}

\maketitle\abstracts{
I present highlights of recent CLEO results on the hadronic decay modes of the
 B meson. Topics covered include two body rare B decays modes, a new 
measurement of $(B^0 \to D^0 K^+) / (B^0 \to D^0 \pi^+ )$, an in depth 
study of the decay $B^0 \to D^* \rho$, as well as a preliminary search for 
charge asymmetry in the decay $B^0 \to K^* \pi$.
}

\section{The CLEO Detector and CLEO Data}
The CLEO detector has had three incarnations. The original CLEOII
 detector, as described in Ref\,\,\cite{CLEOIIpaper}, 
consists of drifts chambers using an 
argon ethane gas mixture, time of flight counters, a CsI calorimeter, and
 muon chambers. It was upgraded to include a 3 layer 2 sided silicon 
vertex detector, and the drift gas in the drift chamber was changed to a 
helium propane mixture to give improved tracking and $dE/dx$ particle 
identification. This configuration was named the CLEOII.5 detector\,\,\cite{CLEOII5paper}. 
A subsequent major overhaul
 included installation of a four layer double sided silicon vertex detector, 
a new drift chamber, and a RICH detector. This configuration is referred to 
as the CLEOIII detector\,\,\cite{cleo3det}. The datasets which are used in the following were 
taken mainly at the $\Upsilon(4S)$ resonance, which decays equally to 
$B^0 \overline{B}^0$ and ${B^+ B^-}$. One third of the data are taken off 
 peak to gain an understanding of continuum backgrounds.
The CLEOII and CLEOII.5 sample comprise some 9 million B pairs, while the 
CLEOIII sample contains approximately 6 million B pairs. 
It is worth noting that the CESR accelerator produced the $\Upsilon(4S)$ at 
rest, allowing for a kinematic constraint on the reconstruction of its decay.

\section{ Rare Two Body Hadronic B Decays }
Rare two body B decays are studied among other reasons because they can 
provide insight into the CKM mixing angle $\gamma$. Their inherent small 
expected Standard Model branching ratios also allow for the possibility
 of detection of new physics due to the interference of  penguin diagrams
 with Standard Model  diagrams. In addition of course, they 
provide us with
 a sample independent of decays to charm final states to test our 
theoretical expectations. The result presented here will comprise an analysis 
using the CLEOIII detector, which is then combined with the CLEOII+CLEOII.5 
samples to produce the final CLEO results on rare two body hadronic 
B decays from a sample of 15 million B pairs.

The branching ratios reported herein result from $B^0$ decays to 
$\pi\pi$, $K\pi$, $KK$, $K^0\pi^0$, $K^0 \overline{K}^0$,$\pi^0\pi^0$, 
$p\overline{p}$ and $\Lambda \overline{\Lambda} $. Also
 included are results from $B^+$ decays to $\pi\pi^0$, $K\pi^0$, 
$\pi K^0$, $K K^0$ and $p \overline{\Lambda}$.

\subsection{The CLEOIII Analysis Method}

The CLEOIII data was analyzed in much the same fashion as that of CLEOII 
and CLEOII.5 - see Ref\,\,\cite{lastkpipaper}. A series of hard, but well understood, quality cuts
were placed on tracks, showers in the calorimeter, $\pi^0$'s and 
$K_s$'s.
Cuts were also placed on the beam constrained B mass, 
$M_B=\sqrt{ E^2_{Beam}- p^2_{Candidate}} $, for the final state in question, 
as well as on $\Delta E = E_{Candidate}-E_{Beam}$ and the cosine 
of the sphericity angle of the candidate. 

For charged $\pi$ and $K$ final states, the particle identification 
likelihoods of the RICH and $dE/dx$ were combined and cut on. The 
cut efficiency was calibrated using data decays  $D^* \to (K\pi)\pi$
and chosen to accept 90\% of real $\pi$'s and $K$'s, while allowing only 
an 11 \% fake rate for $\pi$'s faking $K$'s, and an 8\% fake rate 
for K's faking $\pi$'s. For proton identification, only the RICH 
detector was used, and cuts were calibrated on data decays of $\Lambda\to 
p \pi$, as well as $D^* \to (K\pi)\pi$, resulting in a measured 
76\% (71\%) $p$ ( $\overline{p}$)  efficiency and a 1\% fake rate from charged K's.

The resulting data were then used in an extended maximum likelihood fit using 
$M_B$, $\Delta E$, $\cos{\theta_B}$ and $\cal F$, where the Fisher discriminant, $\cal F$
is a linear combination of the direction of the thrust axis of 
the candidate, 9 Virtual Calorimeter bins\footnote{The Virtual 
Calorimeter consists of a weighted scalar sum of momenta in 10 degree bins 
around the candidate sphericity bins.} and the momenta of the highest 
electron, $\mu$, $K$, and proton in the event\footnote{This 
last term is new to the CLEO3 analysis.}. The signal Fisher shape 
is derived from Monte Carlo studies, and a common background Fisher 
shape is used in all modes. 
 The resulting Fisher shapes give a 1.4 $\sigma$ separation between the 
signal peaks and background peaks.
The event likelihood is formed using each of the 13 modes above, 
as well as cross feeds from other B modes, and a term representing 
the non-B background.

\subsection{Results}
The CLEO III measurements using the likelihood
functions described above are combined with the CLEOII and CLEOII.5 results 
reported in 
Ref\,\,\cite{lastkpipaper} and Ref\,\,\cite{allrareb} and reported in 
Table\,\,\ref{table:cleoAll}. For some 
modes previously unpublished likelihood functions are used. The baryonic modes
and the \ks\ks\ mode were analyzed here with the full CLEO II data set for the 
first time. Full details can be found in Ref\,\,\cite{cleo3rareb}. All results
 are competitive with recent results from the BaBar and Belle collaborations\,\,\cite{babarandbelle}.

\begin{table*}[htbp]
\begin{center}
\caption{
Experimental results for CLEO II, CLEO III, and
both datasets combined.  Significances include systematic errors.
Note that the $p{\bar p}$ analysis in Ref.\,\,$^{5}$
%\cite{allrareb} 
was 
done in only a subset of the full CLEO II dataset, so the 
``combined'' result is simply the CLEO III upper limit.  Upper limits
are 90\% confidence level. CLEO II results are taken from Ref.\,\,$^{5}$
%\cite{allrareb},
except for the $\kz\kzb$, \pL\, and \LL\ modes which were analyzed
in this work with the full CLEO II dataset for the first time. }
\smallskip
\begin{tabular}{|l|cc|cc|cc|}
\hline
 & \multicolumn{2}{|c|}{CLEO II - Ref. 5}
 & \multicolumn{2}{|c|}{CLEO III} 
 & \multicolumn{2}{|c|}{Combined}\\
\hline
Mode & Significance & $ {\cal B}\times 10^{6}$
     & Significance & $ {\cal B}\times 10^{6}$
     & Significance & $ {\cal B}\times 10^{6}$
     \\ \hline\hline
$\pi^+\pi^-$ & 4.2    & 4.3$^{+1.6+0.5}_{-1.4-0.5}$ 
             & 2.6    & 4.8$^{+2.5+0.8}_{-2.2-0.5}$ 
             & 4.4    & 4.5$^{+1.4+0.5}_{-1.2-0.4}$ \\
$\pi^+\pi^0$ & 3.2    & 5.6$^{+2.6+1.7}_{-2.3-1.7}$  
             & 2.1    & 3.4$^{+2.8+0.8}_{-2.0-0.3}$ 
             & 3.5    & 4.6$^{+1.8+0.6}_{-1.6-0.7}$ \\
$\pi^0\pi^0$ & 2.0    & $(<5.7)$                    
             & 1.8    & $(<7.6)$                    
             & 2.5    & $(<4.4)$                    \\ \hline
$K^+\pi^-$   & $12$   & 17.2$^{+2.5+1.2}_{-2.4-1.2}$
             & $ >7$  & 19.5$^{+3.5+2.5}_{-3.7-1.6}$
             & $>7$   & 18.0$^{+2.3+1.2}_{-2.1-0.9}$\\
$\kz\pi^+$   & 7.6    & 18.2$^{+4.6+1.6}_{-4.0-1.6}$
             & 4.6    & 20.5$^{+7.1+3.0}_{-5.9-2.1}$
             & $>7$   & 18.8$^{+3.7+2.1}_{-3.3-1.8}$\\
$K^+\pi^0$   & $6.1$  & 11.6$^{+3.0+1.4}_{-2.7-1.3}$
             & $5.0$  & 13.5$^{+4.0+2.4}_{-3.5-1.5}$
             & $>7$   & 12.9$^{+2.4+1.2}_{-2.2-1.1}$\\
$\kz\pi^0$   & 4.9    & 14.6$^{+5.9+2.4}_{-5.1-3.3}$  
             & 3.8    & 11.0$^{+6.1}_{-4.6}\pm2.5$  
             & 5.0    & 12.8$^{+4.0+1.7}_{-3.3-1.4}$\\ \hline
$K^+K^-$     & -      & $(<1.9)$                    
             & -      & $(<3.0)$                    
             & -      & $(<0.8)$                    \\
$\kz K^-$    & -      & $(<5.1)$                    
             & -      & $(<5.0)$                    
             & -      & $(<3.3)$                    \\
$\kz\kzb$    & -      & $(<6.1)$                    
             & -      & $(<5.2)$                    
             & -      & $(<3.3)$                    \\ \hline
$\ppbar$     & -      & $(<7.0)$                    
             & -      & $(<1.4)$                    
             & -      & $(<1.4)$                    \\
$\pL$        & -      & $(<2.0)$                    
             & -      & $(<3.2)$                    
             & -      & $(<1.5)$                    \\
$\LL$        & -      & $(<1.8)$                    
             & -      & $(<4.2)$                    
             & -      & $(<1.2)$                    \\ \hline
\end{tabular}
\label{table:cleoAll}
\end{center}
\end{table*}

\subsection{The ratio ${ B(B^- \to D^0K^- )/{B(B^- \to D^0\pi^- )}}$}
Using the same technique as described above, we also examined the 
ratio of branching fractions 
of $B^- \to D^0K^-$  to $B^- \to D^0\pi^- $ using the decays
$D^0\to K^-\pi^+$, $D^0\to K^-\pi^+\pi^0$, and $D^0\to K^-\pi^+\pi^-\pi^+$.
 We apply particle identification only to the fast $K$ and $\pi$ 
in the decay and account for the feed-through from the $\pi$ channel 
into the $K$ channel. The ratio allows for a small systematic error 
due mainly to particle identification uncertainties. The result is:
\begin{equation}
{B(B^- \to D^0K^- )\over{B(B^- \to D^0\pi^- )} } = ( 9.9 ^{+1.4 + 0.7}_{-1.2 -0.6} ) \times 10^{-2}
\end{equation}

\section{The Decay $B \to D^*\rho$}
Using the CLEOII and CLEOII.5 data sample, and using an extended maximum likelihood 
fit in the B candidate mass and $\rho$ candidate mass, we report:
\begin{eqnarray}
B( B^- \to D^{*0} \rho^- ) & = & ( 0.98 \pm 0.06 \pm 0.16 \pm 0.05) \% \cr
B( \overline{B}^0 \to D^{*+}\rho^- ) &=& (0.68\pm 0.03 \pm 0.09 \pm 0.02) \%
\end{eqnarray}
where the first systematic error is dominated by $\pi^0$ reconstruction 
errors, and the second is due to $D$ and $D^*$ branching ratio 
uncertainties. From this we can extract the ratio of BSW coupling 
constants\,\,\cite{bsw} $|a_2/a_1|  = 0.21 \pm 0.03 \pm 0.05 \pm 0.04 \pm 0.04$, 
where the last uncertainty is from the error on the ratio of charged 
decays of the $\Upsilon(4S)$ to uncharged decays.

The differential decay rate can be described in terms of the helicity 
amplitudes $H_0$, $H_+$, and $H_-$ using the cosine of the opening 
angles of the $D^*$   and $\rho$  
 decays, and the relative angle  of 
the two decays planes. These amplitudes are extracted from the 
data using an extended maximum likelihood fit whose inputs are 
the masses of the $B$ and $\rho$ candidates, and the angles described 
above. Acceptance corrections are taken into account in the fit.
The results are listed in Table\,\,\ref{table:dstarrho}.

\begin{table*}[htbp]
\begin{center}
\caption{Helicity amplitudes and phases for the decay $B\to D^*\rho$. 
The angles $\alpha_{(+,-)}$ are the phases of the helicity amplitudes 
$|H_{(+,-)}|$ relative to $|H_0|$. The results for the resulting 
relative longitudinal decay rates (or polarization)  are also shown.}  
\smallskip
\begin{tabular}{|c|c|c|}
\hline
                     & $D^{*0}$ & $D^{*+}$ \\ \hline
   $|H_0|$    & $0.944 \pm 0.009 \pm 0.009$ & $0.941 \pm 0.009 \pm 0.006$ \\ \hline
   $|H_+|$    & $0.122\pm0.040\pm0.010$ & $0.107 \pm 0.031 \pm 0.011$ \\ \hline
   $\alpha_+$ & $1.02 \pm 0.28 \pm 0.11$ & $1.42 \pm 0.27 \pm 0.04$ \\ \hline  
   $|H_-|$  & $0.306 \pm 0.030 \pm 0.025$ & $0.322 \pm 0.025 \pm 0.016$ \\ \hline
    $\alpha_-$ & $0.65 \pm 0.16 \pm 0.04$ & $0.31 \pm 0.12 \pm 0.04 $ \\ \hline
    $\Gamma_L / \Gamma$ & $0.892 \pm 0.018 \pm 0.016$ & $0.885 \pm 0.016 \pm 0.012$ \\ \hline
\end{tabular}
\label{table:dstarrho}
\end{center}
\end{table*}

From the amplitudes, one can form the relative longitudinal and 
transverse decay rates. The factorization ansatz\,\,\cite{bsw} predicts that the 
longitudinal polarization of the $D^{*+}\rho^-$ decay should be 
equal to the longitudinal polarization of the semileptonic decay 
$\overline{B}^0 \to D^{*+} l^- \overline{\nu} $ evaluated at  
$q^2 = M^2_\rho$. The results are in agreement with this prediction. 
 Full results are discussed in Ref\,\,\cite{dstarrho}.

\section{ Charge Asymmetry in $B^0 \to K^* \pi$ }
 The investigation of charge asymmetries in neutral B decays can yield
 insights into CP violation. Using the CLEOII and CLEOII.5 data 
sample, we have
investigated the asymmetry ${\cal A}_{CP}$ between the decays $\overline{B}^0 
\to K^{*-}(892)\pi^+$ and  $B^0 \to K^{*+}(892)\pi^-$ by looking 
at the asymmetry ${\cal A}_{+-}$ which is the difference in rates between 
the final states with a positive or negative fast $\pi$. Monte 
Carlo simulations indicate that ${\cal A}_{+-}$ is $99.99\%$ correlated with 
${\cal A}_{CP}$ for this particular decay mode. The final states of interest 
include $(K_s \pi^+) h^-$ and $(K^-\pi^0) h^+$, where $h$ indicates 
that the particle has not undergone particle identification cuts.

The asymmetry is extracted from a maximum likelihood fit using 
the $B$ candidate mass, $\Delta E$, the Dalitz plot, the B decay 
angle, $dE/dx(h^+)$ and a Fisher discriminant composed of the virtual 
calorimeter, the Fox-Wolfram event shape R2, and the cosine of the angle of 
 the candidate thrust axis with respect to the beam. 
Included in the fit are distributions for the 
$K^{*+}(982) \pi^- / K^-$, $K^{*+}(1430) \pi^- / K^-$, $\rho^0 K_s^0$, 
$K_s \pi(\pi/K)$ non resonant decays, as well as $K^{*0}(892) \pi^0$, 
 $K^{*0}(1430) \pi^0$, and $K^-\pi^+\pi^0$ non resonant
 decays. We exclude background slices in the Dalitz plot
 consisted with feed-throughs from the decays of $B$ mesons to $D\pi$, 
$\psi K_s$, and $\psi \pi^0$ final states. In addition to these signal modes, 
generic $B\overline{B}$ decays and continuum feed-through are in 
the likelihood.

The preliminary value extracted for ${\cal A}_{CP}$ is found  to be:
\begin{equation}
  { \cal A}_{CP}(B\to K^{*\pm}(892)\pi^{\mp}) =
       0.26 ^{ +.33+0.10}_{-.34-0.08}.
\end{equation}
By integrating the likelihood function, this leads to a $90\%$ confidence 
level interval  of :
 $ - 0.31 < {\cal A}_{CP}(B\to K^{*\pm}(892)\pi^{\mp} )
                    < 0.78 $.
The dominant systematic uncertainty for this measurement is our lack of 
knowledge of the interference terms between the various signal components. 
This result is the first limit on ${\cal A}_{CP}$ in this mode. Theoretical 
 predictions for this asymmetry range from $-0.19$ to $0.47$ as found in 
 Ref\,\,\cite{acptheory}.
 
\section{Conclusions}
We have presented the final combined CLEO results on rare 2 body B decay modes. 
These results are in agreement with, and still  competitive with, recent 
results from both the BaBar and Belle  collaborations\,\,\cite{babarandbelle}. 
A by product of this analysis has been a new result for the ratio of 
branching ratios 
$ B(B^- \to D^0 K^- )/ B(B^- \to D^0 \pi^- )$ which is again 
competitive with results from other collaborations. 

 We have also presented detailed results on the branching ratios and 
helicity amplitudes of the decay $B\to D^* \rho$, as well as a determination 
of the ratio of BSW coefficients $|a_2/a_1|$, and a confirmation of the 
factorization ansatz. 

 We have also presented the first limits on ${\cal A}_{CP}$ in the decay 
$B\to K^* \pi$. 

 The CLEO collaboration is now moving into a new era,  dubbed the CLEO-c 
era, and will from herein concentrate on low energy $e^+ e^-$ collisions near 
the charm production threshold. 
Although this  represents a new physics sample, it is clear that the 
older $\Upsilon(4S)$ sample collected continues to bear fruit in analysis.

\section*{Acknowledgments}
 We gratefully acknowledge the effort of the CESR staff in providing us with excellent luminosity and  running conditions. This work was supported by the National Science Foundation, the U.S. Department of Energy, the Research Corporation, and the Texas Advanced Research Program. This author in  particular would like to thank the conference  organizers for  providing us with the opportunity to meet in such a inspiring setting.

\end{document}